\documentclass[aps,prl,english,superscriptaddress,reprint,showpacs]{revtex4-1}
\usepackage{amstext,amssymb,graphicx,babel,soul,bm,textcomp,inputenc,upgreek,gensymb}

\usepackage{comment,color}

\begin{document}

\title{Interplay of Dirac nodes and Volkov-Pankratov surface states in compressively strained HgTe}

\author{David~M.~Mahler}
\affiliation{Physikalisches Institut der Universit\"{a}t
W\"{u}rzburg, 97074 W\"{u}rzburg, Germany}
\affiliation{Institute for Topological Insulators, 97074 W\"{u}rzburg, Germany}

\author{Julian-Benedikt~Mayer}
\affiliation{Institut f\"{u}r Theoretische Physik und Astrophysik,
Universit\"{a}t W\"{u}rzburg, 97074 W\"{u}rzburg, Germany}

\author{Philipp~Leubner}
\affiliation{Physikalisches Institut der Universit\"{a}t
W\"{u}rzburg, 97074 W\"{u}rzburg, Germany}
\affiliation{Institute for Topological Insulators, 97074 W\"{u}rzburg, Germany}

\author{Lukas~Lunczer}
\affiliation{Physikalisches Institut der Universit\"{a}t
W\"{u}rzburg, 97074 W\"{u}rzburg, Germany}
\affiliation{Institute for Topological Insulators, 97074 W\"{u}rzburg, Germany}

\author{Domenico~Di~Sante}
\affiliation{Institut f\"{u}r Theoretische Physik und Astrophysik,
Universit\"{a}t W\"{u}rzburg, 97074 W\"{u}rzburg, Germany}

\author{Giorgio~Sangiovanni}
\affiliation{Institut f\"{u}r Theoretische Physik und Astrophysik,
Universit\"{a}t W\"{u}rzburg, 97074 W\"{u}rzburg, Germany}

\author{Ronny~Thomale}
\affiliation{Institut f\"{u}r Theoretische Physik und Astrophysik,
Universit\"{a}t W\"{u}rzburg, 97074 W\"{u}rzburg, Germany}

\author{Ewelina~M.~Hankiewicz}
\affiliation{Institut f\"{u}r Theoretische Physik und Astrophysik,
Universit\"{a}t W\"{u}rzburg, 97074 W\"{u}rzburg, Germany}

\author{Hartmut~Buhmann}
\affiliation{Physikalisches Institut der Universit\"{a}t
W\"{u}rzburg, 97074 W\"{u}rzburg, Germany}
\affiliation{Institute for Topological Insulators, 97074 W\"{u}rzburg, Germany}

\author{Charles~Gould}
\affiliation{Physikalisches Institut der Universit\"{a}t
W\"{u}rzburg, 97074 W\"{u}rzburg, Germany}
\affiliation{Institute for Topological Insulators, 97074 W\"{u}rzburg, Germany}

\author{Laurens~W.~Molenkamp}
\affiliation{Physikalisches Institut der Universit\"{a}t
W\"{u}rzburg, 97074 W\"{u}rzburg, Germany}
\affiliation{Institute for Topological Insulators, 97074 W\"{u}rzburg, Germany}

\begin{abstract}
Abstract inline
\end{abstract}


\maketitle

{\bf Preceded by the discovery of topological insulators, Dirac and Weyl semimetals have become a pivotal direction of research in contemporary condensed matter physics. While easily accessible from a theoretical viewpoint, these topological semimetals pose a serious challenge in terms of experimental synthesis and analysis to allow for their unambiguous identification. In this work, we report on detailed transport experiments on compressively strained HgTe. Due to the superior sample quality in comparison to other topological semimetallic materials, this enables us to resolve the interplay of topological surface states and semimetallic bulk states to an unprecedented degree of precision and complexity. As our gate design allows us to precisely tune the Fermi level at the Weyl and Dirac points, we identify a magnetotransport regime dominated by Weyl/Dirac bulk state conduction for small carrier densities and by topological surface state conduction for larger carrier densities. As such, similar to topological insulators, HgTe provides the archetypical reference for the experimental investigation of topological semimetals.}

The discovery of topological insulators has inspired a remarkably broad interest in materials
whose band structures exhibit relativistic properties.
The effects of a linear dispersion in one-dimensional edge channels
of quantum spin Hall insulators~\cite{konig_quantum_2007}, as well as in
two-dimensional surface states of three-dimensional topological insulators~\cite{hsieh_topological_2008,brune_quantum_2011}, have already been extensively studied.
The implications of a linear band dispersion in three-dimensional conductors, however,
have only recently begun to be explored.
Such materials, dubbed Dirac or Weyl semimetals, represent a condensed matter realization
of the Weyl/Dirac equations, and may provide an environment for studying
the properties of quasiparticles which have been postulated, but not yet unambiguously demonstrated, to exist in nature.

In many of these materials~\cite{armitage_weyl_2017}, the Weyl or Dirac band crossing is caused by a band inversion,
and is intimately connected to the point group symmetry of the crystal lattice. 
This lends similarities to the prototypical setup of topological insulators.
In fact, both in the alkali pnictide (AB$_3$, where A=(Na,K,Rb), B=(As,Sb,Bi)) and Cd$_2$As$_3$ families that boast a number of important Weyl/Dirac compounds, the inversion occurs between
metallic $s$-like and chalcogenic $p$-like orbitals, a situation very similar to that found in HgTe. The correspondence in terms of band structure between these compounds and HgTe has indeed been known since the 1970s~\cite{caron_cdas_1977}.
The common motif is that, for the alkali pnictides and Cd$_2$As$_3$, the $p$-like $j=3/2$ bands ($\Gamma_8$ in the $T_d$ point group) cross and yield Dirac (or Weyl) points, while
in HgTe the $\Gamma_8$ bands just touch, derives from the higher (zincblende) point group symmetry of the HgTe crystal. Small crystal distortions from the zincblende symmetry, as present in Weyl/Dirac semimetals, are sufficient to crucially modify the electronic structure at low energies.

\begin{figure}
\begin{centering}
\includegraphics[width=1\columnwidth]{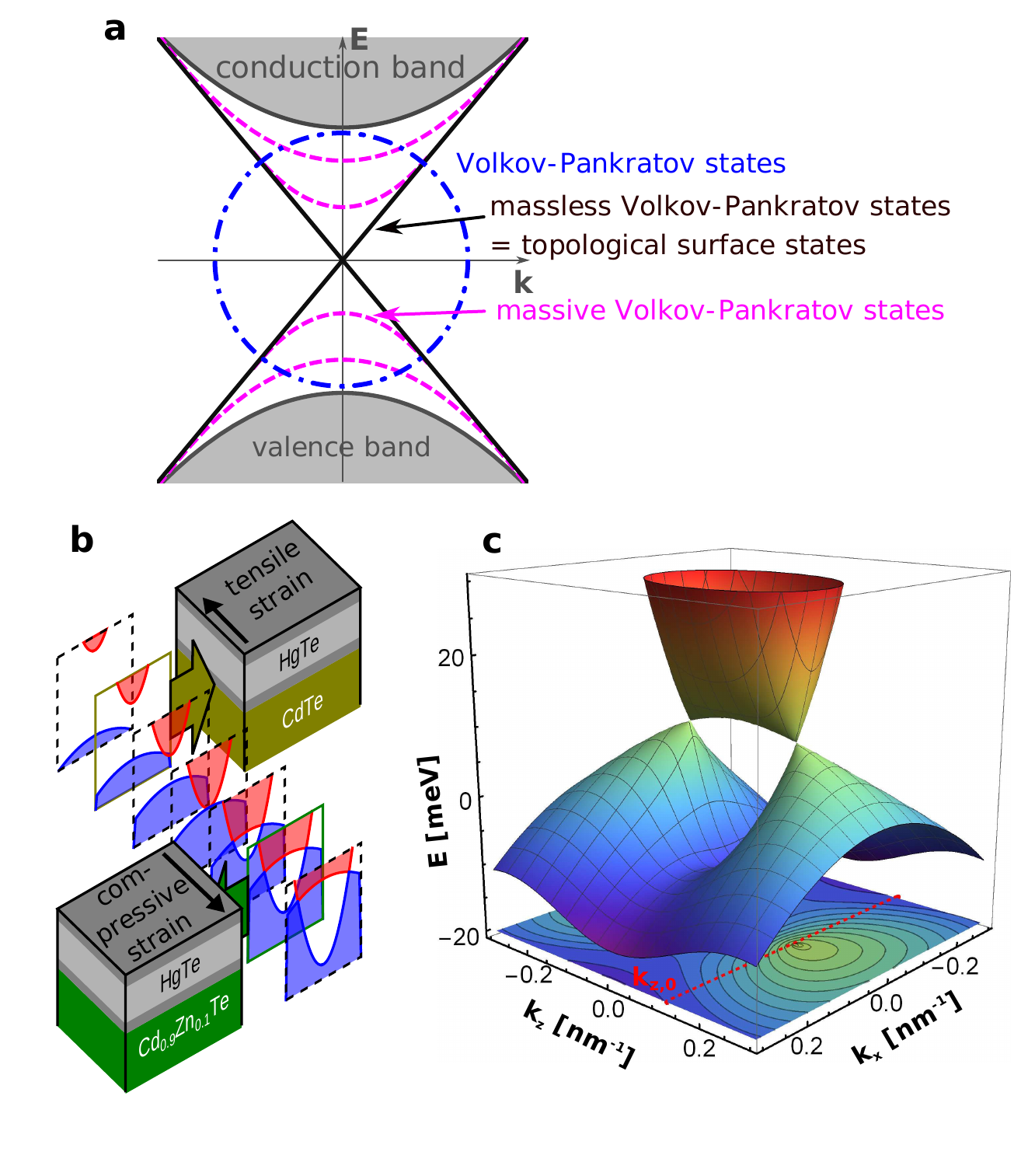}
\par\end{centering}
\caption
{\textbf{a} shows a schematic picture of the band structure at an interface of two semiconductor materials with mutually inverted bulk bands (grey) with two types of interface states (named after the authors of Ref.~\cite{Volkov_two-dimensional_1985} Volkov-Pankratov states), the massless interface states in black and the massive ones in dashed magenta.
\textbf{b} Strain-imposed growth of HgTe. Tensile strain is realized by a CdTe substrate, compressive strain by a $\text{Cd}_{0.9} \text{Zn}_{0.1} \text{Te}$ virtual substrate. Plotted as a function of $k_{z}$, i.e,. along the growth direction, the band structure profile changes from quadratic band touching for unstrained HgTe to a topological bulk gap for tensile strain and a pair of linear level crossing for compressive strain. 
\textbf{c} Band structure plot for the compressive strain regime in the $k_z$-$k_x$ momentum plane.
The bulk inversion asymmetry that would split the Dirac points into Weyl nodes is not accounted for in the calculation. 
}
\label{fig:bandstructure}
\end{figure}

In the 1980s, Volkov and Pankratov~\cite{Volkov_two-dimensional_1985} studied the interface between 
two semiconductor materials with mutually inverted bands, and reported a resulting band structure as 
depicted in Fig.~\ref{fig:bandstructure}\textbf{a}. It includes linear dispersing massless surface
states -the states that are now interpreted as the defining property of topological insulators, and 
topologically trivial massive surface states. While the latter are currently commonly referred to as 
massive Volkov-Pankratov states~\citep{Tchoumakov_2017}, the former are often called 
topological surface states. Here, we use the historically accurate nomenclature of referring 
to both types of states identified in Ref.~\cite{Volkov_two-dimensional_1985} as Volkov-Pankratov states,
and differentiating between them by qualifying them as either massless or massive.
A generic \cite{Yang2014, Yi_2014} even though experimentally often overlooked or neglected implication of bulk band inversion in Dirac or Weyl materials is the concomitant creation of these same Volkov-Pankratov states at energies even significantly away from the bulk crossing point. 
As such, while the bulk band structure of these systems exhibits a three-dimensional linear dispersion relation, massless Volkov-Pankratov states continue to support two-dimensional linearly dispersive bands as well. 
Thus, two and three-dimensional conducting states
co-exist, and care must be taken in transport experiments
to unambiguously assign any feature observed in the conduction profile of the sample to its individual origin.

As an additional challenge for the experimental analysis, Weyl and Dirac materials typically have a high carrier density - as a consequence, they are difficult or sometimes even impossible to gate. Moreover, because proper lithographic methods and thin layer approaches have yet to be developed, primitive contacting (such as needles and conducting glues) and patterning methods are usually employed, giving rise to many potential measurement artifacts resulting from inhomogeneous current distribution, which becomes even further enhanced by the application of magnetic fields. Such effects, sometimes summarized by the expression current jetting
~\cite{arnold_negative_2016, yoshida_transport_1980,
reis_search_2016} are, for instance, known to lead to inaccurate mobility measurements. 
Altogether, such concerns have cast significant doubt on the reliability of many of the early experiments on Dirac and Weyl materials~\cite{armitage_weyl_2017}.

The synoptic view of the beforementioned observations strongly suggests that materials of higher quality and more mature synthesis procedures are indispensable to truly discover the enigmatic Dirac and Weyl semimetallic state. In this paper, we report that compressively strained HgTe is an ideal choice for such an undertaking.
It can be grown with high crystalline quality
by molecular beam epitaxy (MBE), leading to exceptionally low inherent carrier densities.
Furthermore, we can use well established lithographic techniques to precisely define a Hall-bar structure with low resistance alloyed ohmic contacts as well as electrostatic gate electrodes. These good contacts and exact device geometries ensure a well-defined homogeneous current distribution.
Most importantly, the low intrinsic doping, together with the inclusion of a gate, allows us to controllably adjust
the level of the Fermi energy via the carrier density within the band structure, tuning the conductance properties
between surface state and bulk Dirac/Weyl node dominated transport. This allows us to confidently attribute transport characteristics to either their bulk or surface origin.

The low energy dispersion of HgTe is given by two quadratically dispersing $\Gamma_{8}$-bands. For unstrained bulk HgTe, these bands are degenerate at the $\Gamma$-point, as sketched in Fig.~\ref{fig:bandstructure}\textbf{b}.
Under tensile strain, this degeneracy is lifted~\cite{liu_transport_1975} due to lowering of the point group symmetry, and a topological bulk gap opens~\cite{brune_quantum_2011, brune_dirac-screening_2014}. The remaining surface conduction stems from the massless Volkov-Pankratov states implied by inversion of the $\Gamma_{8}$-bands with the $\Gamma_{6}$ band, the latter of which, for the unstrained case, is located deep below the Fermi level at $\Gamma$.

Under compressive strain, the $\Gamma_{8}$ degeneracy is likewise lifted, now with the two $\Gamma_{8}$-bands shifting in opposite direction as for tensile strain. This leads to the formation of linear crossing points in the band structure.
The in-plane compressive and tensile strain dependence of the bands around $\Gamma$ is visualized in Fig~\ref{fig:bandstructure}\textbf{b}, where the red domain highlights the conduction and the blue domain the valence regime.  
Experimentally, the tensiley strained 3D TI phase is accomplished by growth
on a CdTe substrate (Fig.~\ref{fig:bandstructure}\textbf{b}). While there is no commercially available substrate with a lattice constant slightly below that of HgTe, compressive strain is experimentally still accessible through a superlattice virtual substrate~\cite{leubner_strain_2016} as sketched in Fig.~\ref{fig:bandstructure}\textbf{b}.	
The linear crossing points are further demonstrated in  Fig.~\ref{fig:bandstructure}\textbf{c} via a band structure plot along the $k_z$-$k_x$ plane in momentum space. It is the result of a 6-band k$\cdot$p calculation of the low energy band structure for the compressively strained case, where the $z$ axis denotes the layer growth direction and the $x,y$ axes the in-plane coordinates.  

\begin{figure}
\begin{centering}
\includegraphics[width=1\columnwidth]{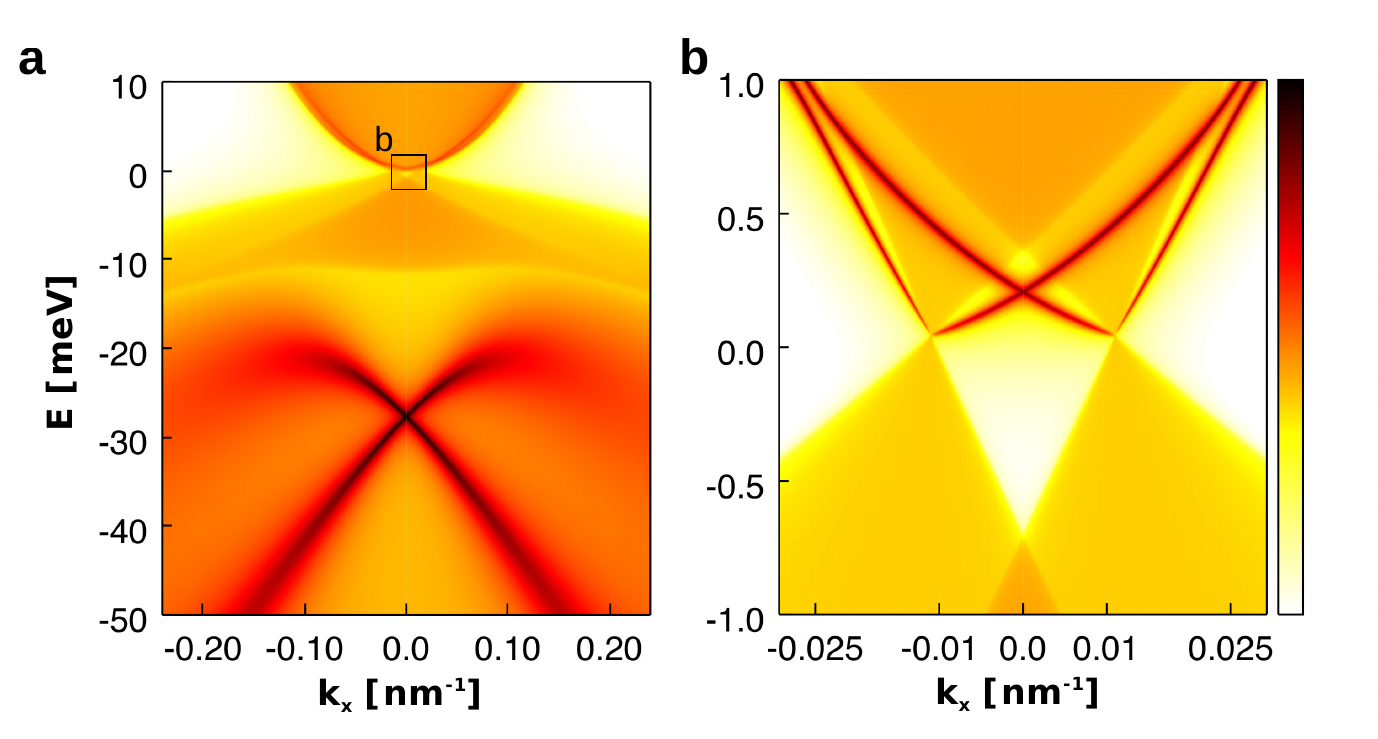}
\par\end{centering}
\caption
{
\textbf{a} Band structure along the $k_x$ direction of the surface Brillouin zone of a semi-infinite thick slab of compressively strained HgTe ($\approx 0.3 \%$) from density functional theory. 
Spectral features in dark red show prominent surface character, while lighter colours display the continuum of bulk states with relatively weaker projections onto the (001) surface. The sharp dark lines highlight the massless Volkov-Pankratov states and Dirac point stemming from the $\Gamma_6$-$\Gamma_8$ band inversion. 
Due to the sharp interface and the absence of a Hartree potential in this calculations, the massive Pankratov states are shifted to high energies and not visible here.
\textbf{b} Zoom of panel {\bf a} around the Fermi level showing two Weyl points with the same chirality $+2$. 
At large momenta, i.e. far from the Weyl points low-energy physics, the surface states originating from the Weyl nodes are the massless Volkov-Pankratov states that disperse in the unoccupied part of the spectrum (cf. the surface states above the Fermi level in panel \textbf{a}). 
The color code identifies the spectral function $A({\text k}_x,\omega)=(-1/\pi){\text Im}G_{\text{surf}}({\text k}_x,\omega)$, where $G_{\text{surf}}({\text k}_x,\omega)$ is the momentum and energy resolved surface Green's function.
}
\label{fig:DFT}	
\end{figure}

More information on the relation between bulk Dirac nodes and the inversion-induced surface states can be obtained from DFT calculations on a semi-infinite thick slab with a tellurium terminated interface to vacuum of compressively strained HgTe at a realistic value of the in-plane strain
(see also supplementary information \cite{sup}). Fig.~\ref{fig:DFT}\textbf{a} shows the calculated slab dispersion for an extended energy range centered around the Fermi level, exhibiting sharp dispersive surface features 20 meV below and right above the Fermi level, which is related to the inversion of the $\Gamma_6$ and $\Gamma_8$ bands. The breaking of spatial inversion symmetry in the zincblende structure splits each Dirac node into four Weyl points~\cite{ruan_symmetry-protected_2016}. For HgTe, Weyl points of opposite chirality project pairwise onto the (001) surfaces connected by inversion. Starting from two Dirac nodes in the unstrained zincblende structure, this yields four Weyl points per surface, with a Berry flux monopole charge of $\pm 2$.  Fig.~\ref{fig:DFT}\textbf{b} highlights the low-energy dispersion for the two surface-projected Weyl points of chirality +2. Weyl points of opposite chirality are connected by a Fermi arc \cite{wan_topological_2011}.
 The separation in momentum space between the Weyl nodes is estimated to be $\sim 0.02$ nm$^{-1}$. As a consequence, zero-field experiments do not offer sufficient resolution to resolve the small energy and momentum scales at hand. As such, we hereafter consider all bulk transport in this sample to be described by Dirac physics and thus will refer to the energy of the crossing points as the bulk Dirac node. From this perspective, the band structure of compressively strained HgTe is virtually identical to that of typical 3D Dirac semimetals such as Cd$_3$As$_2$. 

We report experimental results on a  $66\,\text{nm}$ thick, compressively strained HgTe layer, grown on a virtual substrate consisting of a CdTe/ZnTe multilayer produced by a combination of MBE and atomic layer epitaxy (ALE) on a Si-doped GaAs substrate~\cite{leubner_strain_2016}. This CdTe/ZnTe multilayer has a lattice constant between that of CdTe and ZnTe, and can be exactly tuned by setting the Cd/Zn ratio.
In the present case, a lattice constant of $0.6442\,\text{nm}$ is used to impose a compressive strain
of $\approx 0.3\,\% $ on the HgTe layer (for details see the supplementary information \cite{sup}). To increase sample quality and carrier mobility, two $10\,\text{nm}$ thick protective layers of $\text{Cd}_{0.7}\text{Hg}_{0.3}\text{Te}$ are grown below and on top of the HgTe layer.

\begin{figure}
\begin{centering}
\includegraphics[width=0.9\columnwidth]{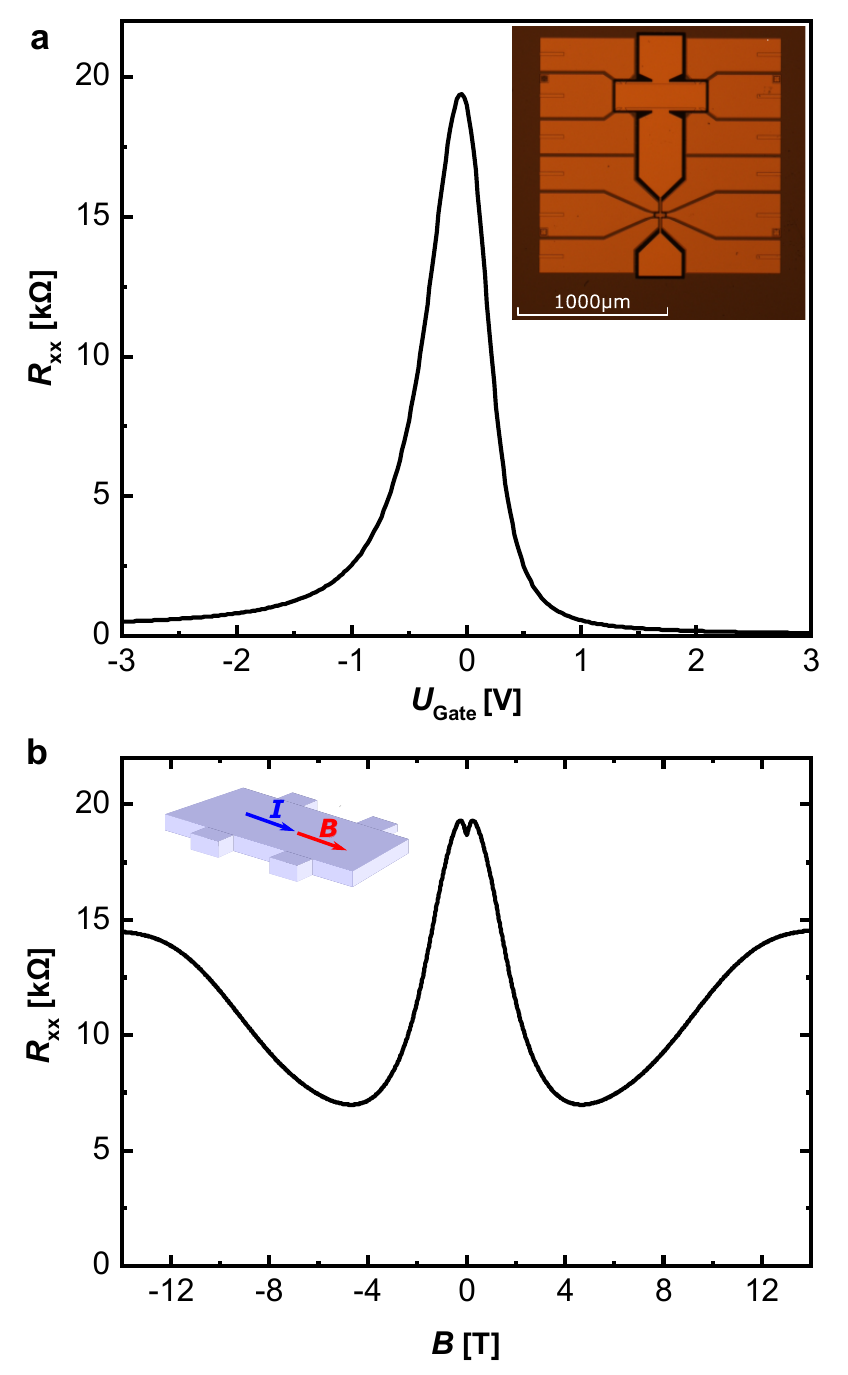}
\par\end{centering}
\caption
{\textbf{a} Longitudinal resistance $R_{xx}$ as a function of the applied gate voltage $U_\text{gate}$ for $B=0$. The inset of \textbf{a} shows an optical microscope picture of the finished sample. \textbf{b} $R_{xx}$ as a function of the applied magnetic field $B$ along the current direction, as indicated by the sketch in the inset. 
}
\label{fig:NMR}
\end{figure}

We pattern the sample using our standard Hall bar mask and optical lithography process into devices such as the one showed in the inset of Fig.~\ref{fig:NMR}. The mask contains two sizes of Hall bars, a larger one with a mesa having a width of $200\,\upmu\text{m}$ and a separation of the longitudinal voltage leads of $600\,\upmu\text{m}$, and a smaller one with a width of $10\,\upmu\text{m}$ and a contact separation of $30\,\upmu\text{m}$. The HgTe mesa's are covered with a $110\,\text{nm}$ thick $\text{SiO}_2 / \text{Si}_3\text{N}_4$ insulator followed by a $100\,\text{nm}$ thick Au gate electrode on top of a $5\,\text{nm}$ Ti sticking layer.
Contacts are fabricated by first using a short dry etching step to provide a clean oxide-free surface for contacting, followed by in situ electron beam evaporation of $50\,\text{nm}$ AuGe and $50\,\text{nm}$ Au. Two separate devices, each containing Hall bars of both sizes were investigated in this study, all yielding consistent results with no substantial discrepancy between either the different devices or sample sizes.
All measurements are carried out using standard low-noise and low-frequency AC techniques, and unless otherwise noted, are done at $2\,\text{K}$.

We first confirm that we can indeed efficiently adjust the carrier density in our sample as demonstrated by the influence of the gate voltage on the zero field longitudinal resistance $R_{xx}$ shown in
Fig.~\ref{fig:NMR}\textbf{a}. The longitudinal resistance changes by three
orders of magnitude, from $R_{xx,\text{max}} = 19.3\,\text{k}\Omega$ around
$U_\text{gate} = 0\,\text{V}$ to $R_{xx}(3\,\text{V})= 95\,\Omega$. The associated gate induced change of
carrier density in the sample is determined by standard Hall measurements.
We find that the density can be tuned from
$-10 \times 10^{11}\,\text{cm}^{-2}$  (p-type) for $-3\,\text{V}$ up to
$10 \times 10^{11}\,\text{cm}^{-2}$ (n-type) for $3\,\text{V}$. The maximum of $R_{xx}$
coincides with the lowest total density as well as  with the change in the carrier type from electron to primarily hole transport.
From this observation we infer that we can precisely tune the Fermi energy to the bulk Dirac node. Given that our Hall bars consist of 3 squares, the maximum corresponds to a sheet resistivity value of approximately  $6.4\,\text{k}\Omega / \square$, and thus of the order of magnitude expected for a diffusive Dirac system.\cite{sbierski_weyl_2014}

Tuning to minimal carrier density, and consequently moving the Fermi energy to the bulk Dirac node level, the longitudinal resistance $R_{xx}$ versus a magnetic field $B$ applied parallel to the
current $I$ is shown in Fig.~\ref{fig:NMR}\textbf{b}.
A significant dip is observed as a function of $B$, corresponding to a reduction of up to $\approx 60\,\%$ of the value of $R_{xx}$ at $B=0$. Such a negative magnetoresistance contribution is a defining feature implied by the chiral anomaly. Originally conceived as a symmetry violation in quantum field theory in comparison to its classical analogue, the chiral anomaly was first discussed in the context of solid-state systems by Nielsen
\textit{et. al.}~\cite{nielsen_adler-bell-jackiw_1983}. There, the emergence of a chiral charge, i.e., an imbalance between left- and right moving Dirac particles, is naturally interpreted to be implied by an external field, as both particle branches are not independent, but connected through the crystal band structure. 
Weyl nodes (and magnetic field-split Dirac nodes) can act as magnetic monopoles
in momentum space due to their Berry curvature, with a magnetic charge given by the chirality~\cite{hosur_recent_2013}.
A magnetic field parallel to the driving electric field causes a pair of Weyl nodes
with different chirality to shift in energy with respect to each other, causing a
redistribution of carriers among the nodes. This increases the longitudinal conductivity $\sigma_{xx}$ upon increasing magnetic field strength. A Boltzmann equation calculation yields $\sigma_{xx} \propto B^{2}$~\cite{son_chiral_2013,burkov_negative_2015}.

Experimentally, the negative magnetoresistance due to the chiral anomaly
contribution is, for certain ranges of magnetic field, often overshadowed by other effects.
For small magnetic fields $B$, a minor increase of $R_{xx}$ is observed, which we attribute to weak anti-localization
based on its field and temperature dependence~\cite{lu_weak_2015, zhang_signatures_2016}. For large magnetic fields (above $\approx 6\,$T), the chiral anomaly contribution becomes overcompensated to yield a total increase of longitudinal resistance, possibly due to impurity-imposed classical mechanisms of magnetotransport \cite{Modern_Magnetic_Materials_2000}.

\begin{figure}
\begin{centering}
\includegraphics[width=1\columnwidth]{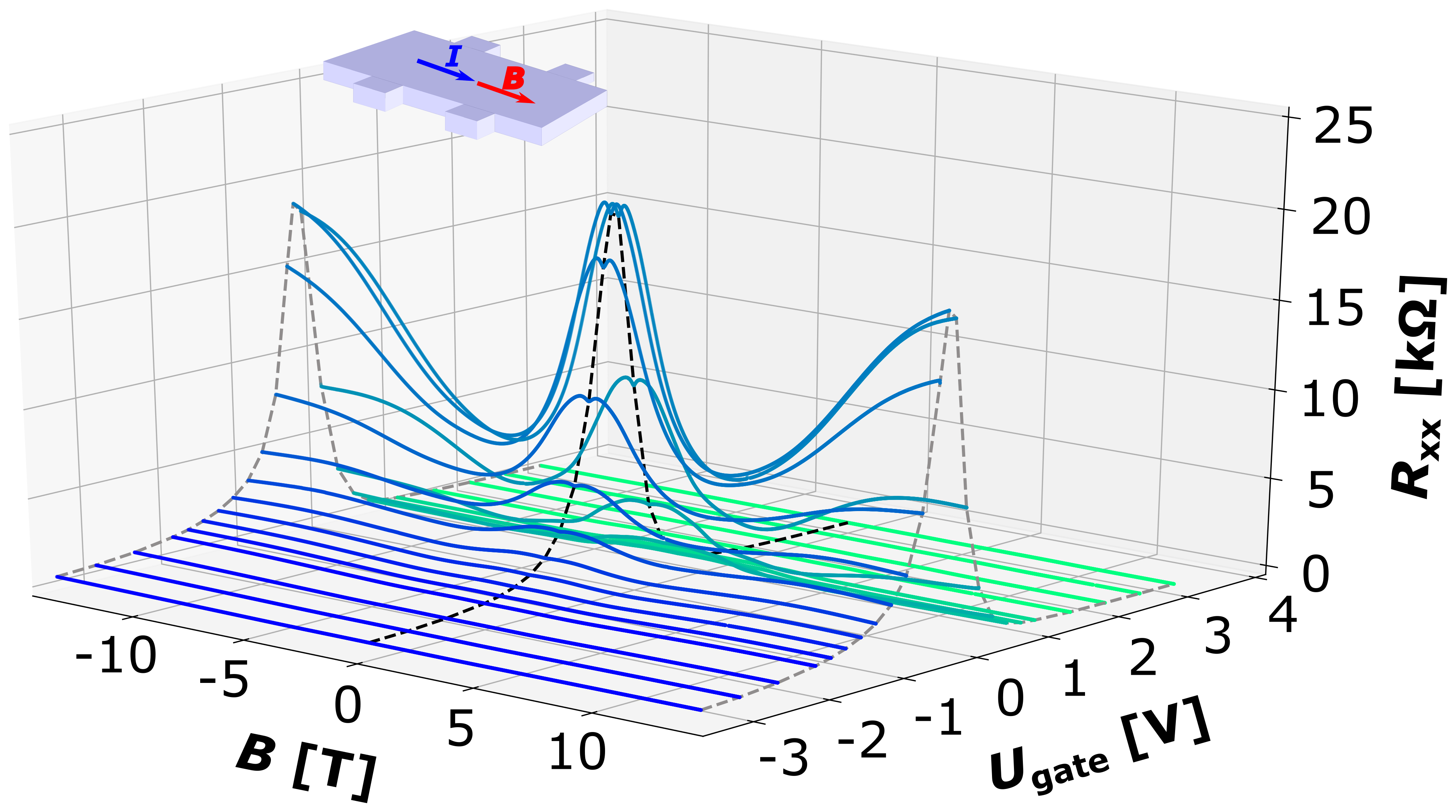}
\par\end{centering}
\caption
{
$R_{xx}$ as a function of the magnetic field applied parallel to the current
for different gate voltages. The dashed lines represent $R_{xx}$ as a function of
the applied gate voltage for zero (black) and maximum magnetic field ($\pm 14\,\text{T}$) (grey).
}
\label{fig:3D}
\end{figure}

Strong evidence connecting the negative magnetoresistance phenomenon in Fig.~\ref{fig:NMR}\textbf{b} to the bulk Dirac nodes in the HgTe band structure derives from the gate voltage dependence of the magnetoresistance data, as presented in Fig.~\ref{fig:3D}.
The data of Figs.~\ref{fig:NMR}\textbf{a} and \ref{fig:NMR}\textbf{b} are included as the $U_\text{gate}=0\,\text{V}$ line and dashed black lines, respectively.
The visualization unambiguously demonstrates that the negative magnetoresistance is strongest, both in absolute numbers and percentage-wise, at $U_\text{gate} \approx 0\,\text{V}$, corresponding to a Fermi energy close to the bulk Dirac nodes.
When the carrier density is increased, the longitudinal resistance $R_{xx}$ at zero field, as well as the magnitude of the negative magnetoresistance, reduce quickly in magnitude.
The reduction of the negative magnetoresistance phenomenon with high absolute gate voltage is equivalently observed for both positive $U_\text{gate}$ (electron transport) and negative $U_\text{gate}$ (hole transport).
A slight asymmetry in the negative magnetoresistance decay for positive versus negative gate voltages 
can be attributed to an asymmetry in the electron and hole mobilities.

\begin{figure}
\begin{centering}
\includegraphics[width=0.9\columnwidth]{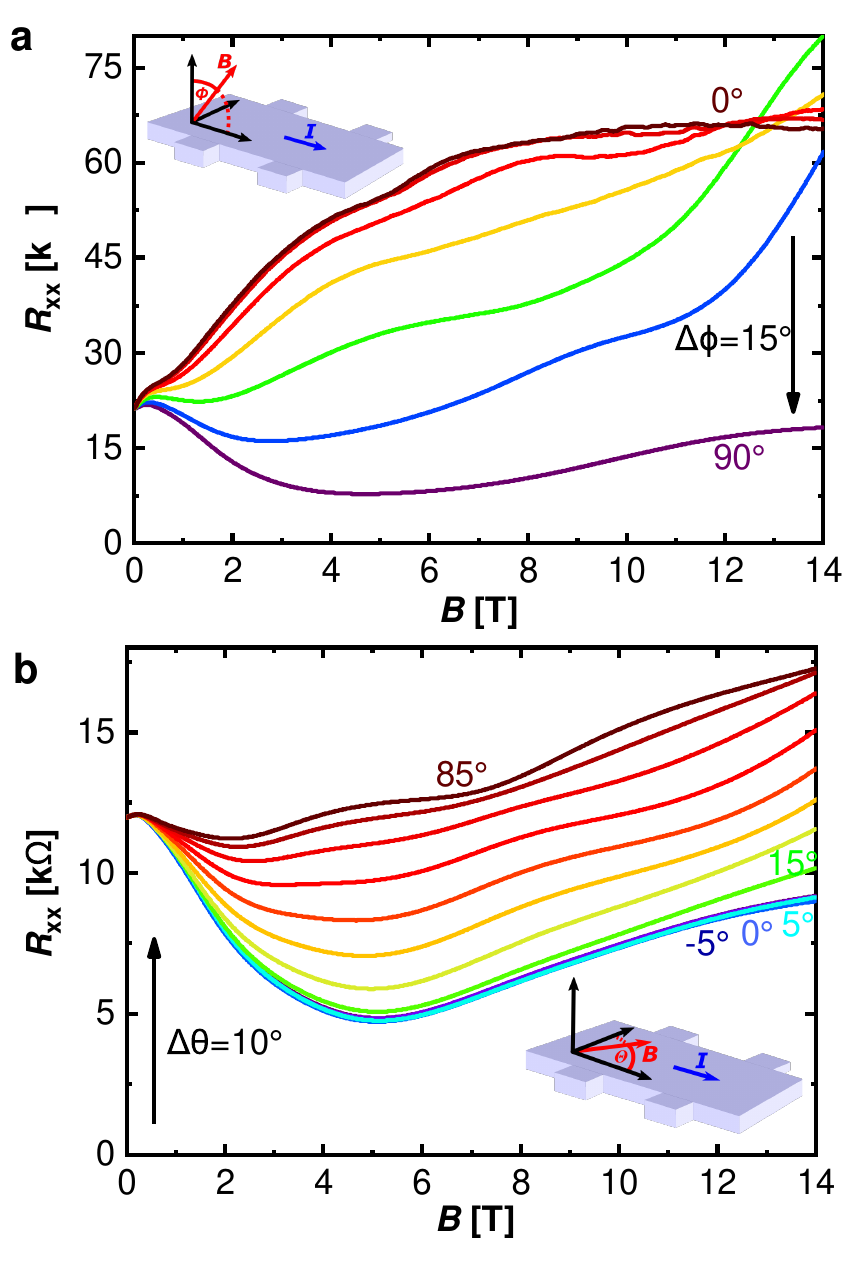}
\par\end{centering}
\caption
{\textbf{a} $R_{xx}$ as a function of the magnetic field $B$ rotated
along $\phi$ from perpendicular to the sample plane towards parallel to the current. In \textbf{b} $R_{xx}$ as a function of in-plane rotation with angle~$\theta$, where $(\theta = 0^{\circ} )$
is parallel to the current and $\theta = 85\degree$ nearly perpendicular to the current (inset) at
$T=0.3\,\text{K}$.
}
\label{fig:Angledependence}	
\end{figure}

A further prototypical feature of the chiral anomaly is the implied angular dependence on the
magnetic field $B$, as only the magnetic field component $B_\parallel$ parallel to the driving electric field $E$
produces an additional current.
The angle dependence for $U_{\text{gate}}(R_{xx,\text{max}})$ is presented
in Fig.~\ref{fig:Angledependence}.
In Fig.~\ref{fig:Angledependence}\textbf{a}
the magnetic field is rotated along the polar angle with $B$ for $(\Phi = 0\degree)$ normal to the plane, and $(\Phi = 90\degree)$ corresponding to $B \parallel E$.
Fig.~\ref{fig:Angledependence}\textbf{b} shows the $B$ dependence of $R_{xx}$
under variation of the azimuthal angle $\theta$ (in the sample plane). The magnetic field direction
is varied from $\theta = 0\degree$, representing $B \parallel E$,
to $B$ nearly $\perp I$ for $\theta = 85\degree$.

While at higher fields, beyond which the chiral anomaly contribution has saturated, classical positive magnetoresistance contributions take over, both parts of Fig.~\ref{fig:Angledependence} show that the amplitude of the negative magnetoresistance depends only on the component of $B$ along the current direction, as expected for the chiral anomaly.
To further confirm the origin of the negative magnetoresistance, a control experiment, on an otherwise identical sample, but with a tensiley strained HgTe layer having a topological insulator band structure, was performed. In that case, only positive magnetoresistance contributions are observed.

We thus conclude that the magnetic field strength and angle
as well as gate voltage dependence of the negative magnetoresistance phenomenon discussed so far is fully
consistent with the expected behavior driven by a chiral anomaly scenario imposed on left- and right moving linearly dispersing branches, which comprises conclusive evidence
for the existence of Dirac nodes in the bulk band structure of our compressively strained HgTe layer.

As already noted, however, the existence of bulk Dirac nodes in no way
precludes the existence of other transport channels. To the contrary, an inverted system with bulk Dirac nodes is generically accompanied by the massless Volkov-Pankratov states. Whether these surfaces contribute to the conductance of the device
depends on the location of the Fermi level and the bending of the gate voltage-induced potential over the device~\cite{brune_dirac-screening_2014}, and the overall quality of the sample material. The gateability of our sample allows us to explore this coexistence.

\begin{figure}
\begin{centering}
\includegraphics[width=1\columnwidth]{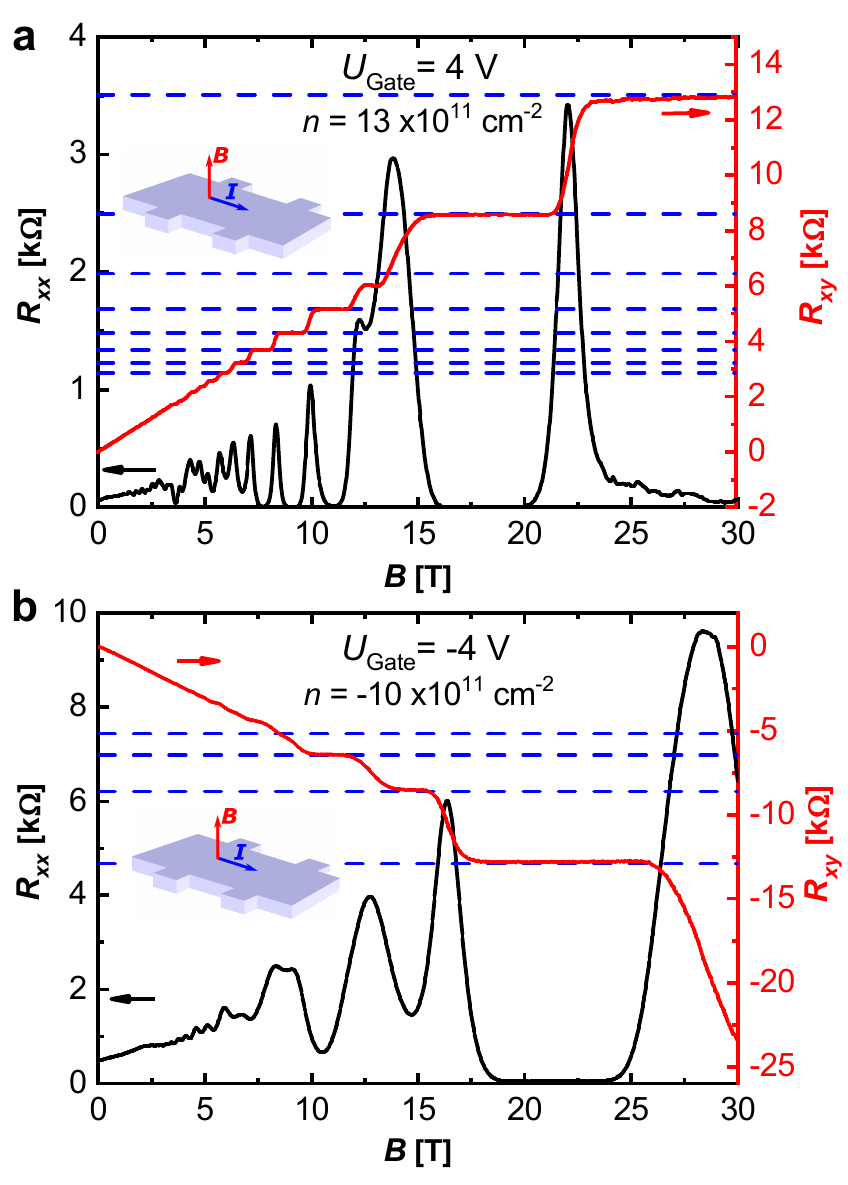}
\par\end{centering}
\caption
{Longitudinal resistance $R_{xx}$ and Hall resistance $R_{xy}$ as a function of the out of plane
magnetic field $B$ for high \textbf{a} electron and \textbf{b} hole densities at $T=0.3\,\text{K}$.
}
\label{fig:Outofplane}
\end{figure}

Our devices show sharply distinct transport behavior when the gate is used to introduce additional carriers. 
This is best observed for measurements in a perpendicular magnetic field.
For illustration, Fig.~\ref{fig:Outofplane} depicts transport data at gate voltages of $\pm 4$V corresponding to highly n-type and highly p-type.
The curves show Shubnikov-de Haas (SdH) oscillations in $R_{xx}$ together with quantum Hall (QH) plateaus.
QH plateaus only exist in two-dimensional systems.
Consistently, the maxima of the SdH-oscillations coincide with the transitions between
QH plateaus, suggesting that also the longitudinal resistance $R_{xx}$ is driven by the same two-dimensional transport channel. Since the investigated sample is a three-dimensional bulk piece,
the two-dimensional character points towards transport carried by a surface state.
For this subset of transport contribution, the p-conducting regime (Fig.~\ref{fig:Outofplane}\textbf{b})
differs from the n-conducting regime (Fig.~\ref{fig:Outofplane}\textbf{a}) mainly
by overall lower mobility, leading to Landau level
broadening, and consequently less pronounced QH plateaus. Mobilities
of $\mu\approx200\,000\,\text{cm}^{-2}/\text{Vs}$ are observed for electrons and
$\mu\approx30\,000\,\text{cm}^{-2}/\text{Vs}$ for holes. These numbers are comparable to the ones reported for
tensiley strained HgTe~\cite{Jost09032017}, which is a topological insulator \cite{brune_dirac-screening_2014}.
The accurate quantization of the plateau levels, i.e., exactly equal to the von Klitzing constant to within the
experimental accuracy of about $1\,\%$, highlights that for these gate voltages,
where the surface states are highly populated, the conduction is dominated by surface transport,
and bulk Dirac contributions are no longer significant.

\begin{figure*}
\begin{centering}
\includegraphics[width=1\textwidth]{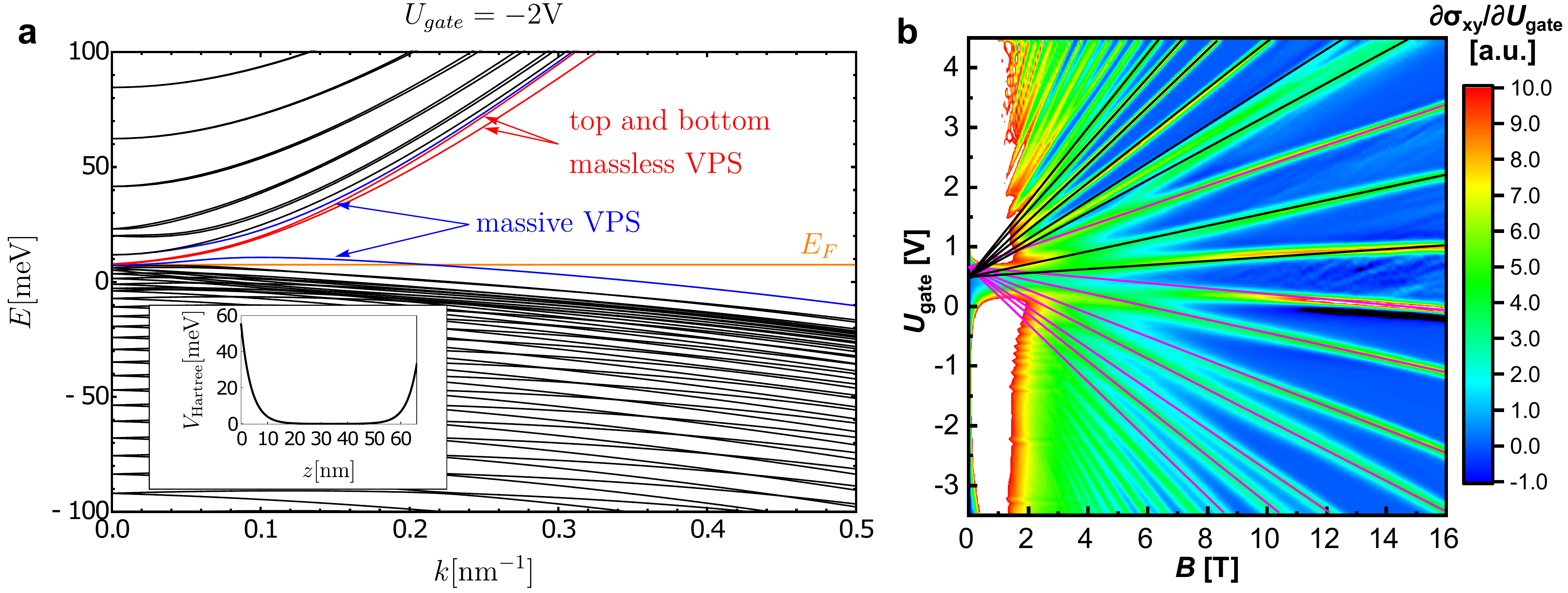}
\par\end{centering}
\caption
{\textbf{a} Calculated k$\cdot$p band structure for a $66\,$nm thick sample including bulk inversion asymmetry terms and a Hartree potential corresponding to an applied gate voltage of $-2\,$V ($n=-4\times 10^{11}\,$cm$^{-2}$). 
The Hartree potential is presented in the inset of $\textbf{a}$. The massless Volkov-Pankratov states are shown in red, while the the massive Volkov-Pankratov states are in blue. 
The bulk subbands are shown in black and the Fermi energy $E_F$ is indicated by the orange line.
\textbf{b} Derivative of the Hall conductivity with respect to the gate voltage ($\partial \sigma_\text{xy}/\partial U_\text{gate}$) as a function of the applied gate voltage $U_\text{gate}$ and magnetic field $B$ at $T=0.02\,\text{K}$. 
The solid lines assign the experimental data to two types of Volkov-Pankratov surface states: the massless ones (black) and massive ones (magenta). 
}
\label{fig:FanChart}
\end{figure*}

An even clearer picture of the interplay between the bulk and surface sources of conduction contributions emerges from
the color scale plot of the gate-voltage derivative of the Hall-conductivity as
a function of gate voltage and magnetic field, presented in Fig.~\ref{fig:FanChart}\textbf{b}.
The derivative values of Hall conductivity (designated in color range from green to red) represent the Landau level (LL) dispersion
of a two-dimensional system with respect to the magnetic field $B$ and the gate voltage $U_\text{gate}$.
For holes ($U_\text{gate}<0\,\text{V}$) a regular pattern of Landau levels is observed.
The splitting between two subsequent Landau levels iterates with smaller and bigger gap.
A zero quantum Hall index is observed for low carrier densities ($U_\text{gate} \gtrsim 0\,\text{V}$)
between two nearly non-dispersive Landau levels, separating the hole and electron regime.
The electron transport regime for higher gate voltages ($U_\text{gate}>0\,\text{V}$)
generally shows a regular pattern of Landau levels. The only exception is the "crossing"
of two Landau levels where the quantum Hall index with $\nu = 4$ would be expected.
This effect can be ascribed to the overlap of two types of surface state LL fans; the ubiquitous
massless Volkov-Pankratov surface states, and the massive Volkov-Pankratov states that arise from 
the high electric field across the HgTe/(Hg,Cd)Te interfaces~\cite{Volkov_two-dimensional_1985} and 
which were recently identified in HgTe-based topological insulators~\cite{Inhofer_2017, Tchoumakov_2017}.
A qualitative description is motivated by 6 $\times$ 6 k$\cdot$p calculations ($\Gamma_8$ and $\Gamma_6$ bands) with hard-wall boundary conditions in the growth direction, 
including compressive strain and a bulk inversion asymmetry term from DFT calculations, as well as a Hartree potential as in Ref.~\cite{brune_dirac-screening_2014}. 
Fig.~\ref{fig:FanChart}\textbf{a} shows the band structure of a $66\,$nm thick sample with an applied Hartree potential corresponding to a gate voltage of $-2\,$V ($n=-4\times 10^{11}$cm$^{-2}$). 
Since the gate voltage is applied from the top surface of the system, such a Hartree potential (shown in the inset of Fig.~\ref{fig:FanChart}\textbf{a}) additionally breaks inversion symmetry. The energies of the massless Volkov-Pankratov states (red) of the top and bottom surface therefore split. 
Additionally, massive Volkov-Pankratov states (blue) form due to the Hartree potential which confines the bulk states, as also discussed in Ref.~\cite{Tchoumakov_2017}. 
For negative gate voltage, the hole-like massive Volkov-Pankratov state crosses the Fermi energy (orange) and thus has the most significant contribution to transport properties at this gate voltage. 
For positive gate voltage, the massless Volkov-Pankratov states dominate the transport behavior since the density of the massive Volkov-Pankratov states is negligible. 
These calculations allow us to assign the black (magenta) Landau levels in Fig.~\ref{fig:FanChart}\textbf{b} to massless (massive) Volkov-Pankratov states.
To show the evolution of the band structure under gate voltage, we provide additional calculations for $U_\text{gate}=0,+2\,$V for the $66\,$nm thick sample in the supplementary material. 
k$\cdot$p calculations and experimental data for a $120\,$nm thick sample are presented along with the analogous analysis in the 
supplementary information \cite{sup}. 

From the above detailed analysis, we conclude that our samples display two distinct transport regimes. First, a narrow gate voltage window around the
resistance maximum at $U_\text{gate} \approx 0\,\text{V}$, where chiral anomaly driven negative magnetoresistance is observed (light
blue traces in Fig~\ref{fig:3D}. This effect is only expected from odd-dimensional Dirac cones, in our case the three-dimensional bulk Dirac cones.
Second, at higher gate voltages (i.e. for finite/high carrier densities) a two-dimensional transport regime is identified by an emerging quantum Hall effect due to the topological surface state of the band inverted material, slightly modified by topologically trivial (massive) Volkov-Pankratov surface states of the material~\cite{Volkov_two-dimensional_1985}.
Our observations provide a simple explanation for the recent findings about a quantum Hall effect in Cd$_2$As$_3$ layers~\cite{Zhang_2017, Schuhmann_2018} and make it evident that extreme care is needed in claiming any contributions from Fermi arcs in the transport properties of Weyl semimetals in general.

To summarize, compressively strained epitaxial HgTe layers have proven to constitute an ideal platform for controlled and reliable transport experiments on
a Weyl/Dirac semimetal. Our experiments emphasize the crucial role played by the inversion-induced 
massless Volkov-Pankratov surface states in this class of topological materials.
As supported by our recent results in this direction, compressively strained HgTe naturally suggest themselves to be an intriguing playground for imposing superconducting proximity effect, where it should be worthwhile probing the superconducting pairing mechanism at Dirac and Weyl nodes.

The work was supported by the DFG (SFB~1170~-~ID:~258499086, Leibniz, 
and \textit{ct.qmat} EXC2147~–~ID:~39085490), the EU (ERC-StG-TOCOTRONICS-Thomale-336012 + ERC-Adv-3TOP-Molenkamp-267436), and the Bavarian ministry of education (ENB and ITI). We acknowledge C.~Br\"{u}ne for useful discussions and F.~Schmidt for help with the measurements. The authors gratefully acknowledge the Gauss Centre for Supercomputing e.V. for providing computing time on the GCS Supercomputer SuperMUC at Leibniz Supercomputing Centre (LRZ). Some measurements at high magnetic fields were performed at the HFML, Nijmegen, the Netherlands.  

\bibliographystyle{apsrev4-1}
\bibliography{_Weyl_Paper_References}

\end{document}


\title{Supplementary Material \protect\\Interplay of Dirac nodes and Volkov-Pankratov surface states in compressively strained HgTe}

\author{David~M.~Mahler}
\affiliation{Physikalisches Institut der Universit\"{a}t
W\"{u}rzburg, 97074 W\"{u}rzburg, Germany}
\affiliation{Institute for Topological Insulators, 97074 W\"{u}rzburg, Germany}

\author{Julian-Benedikt~Mayer}
\affiliation{Institut f\"{u}r Theoretische Physik und Astrophysik,
Universit\"{a}t W\"{u}rzburg, 97074 W\"{u}rzburg, Germany}

\author{Philipp~Leubner}
\affiliation{Physikalisches Institut der Universit\"{a}t
W\"{u}rzburg, 97074 W\"{u}rzburg, Germany}
\affiliation{Institute for Topological Insulators, 97074 W\"{u}rzburg, Germany}

\author{Lukas~Lunczer}
\affiliation{Physikalisches Institut der Universit\"{a}t
W\"{u}rzburg, 97074 W\"{u}rzburg, Germany}
\affiliation{Institute for Topological Insulators, 97074 W\"{u}rzburg, Germany}

\author{Domenico~Di~Sante}
\affiliation{Institut f\"{u}r Theoretische Physik und Astrophysik,
Universit\"{a}t W\"{u}rzburg, 97074 W\"{u}rzburg, Germany}

\author{Giorgio~Sangiovanni}
\affiliation{Institut f\"{u}r Theoretische Physik und Astrophysik,
Universit\"{a}t W\"{u}rzburg, 97074 W\"{u}rzburg, Germany}

\author{Ronny~Thomale}
\affiliation{Institut f\"{u}r Theoretische Physik und Astrophysik,
Universit\"{a}t W\"{u}rzburg, 97074 W\"{u}rzburg, Germany}

\author{Ewelina~M.~Hankiewicz}
\affiliation{Institut f\"{u}r Theoretische Physik und Astrophysik,
Universit\"{a}t W\"{u}rzburg, 97074 W\"{u}rzburg, Germany}

\author{Hartmut~Buhmann}
\affiliation{Physikalisches Institut der Universit\"{a}t
W\"{u}rzburg, 97074 W\"{u}rzburg, Germany}
\affiliation{Institute for Topological Insulators, 97074 W\"{u}rzburg, Germany}

\author{Charles~Gould}
\affiliation{Physikalisches Institut der Universit\"{a}t
W\"{u}rzburg, 97074 W\"{u}rzburg, Germany}
\affiliation{Institute for Topological Insulators, 97074 W\"{u}rzburg, Germany}

\author{Laurens~W.~Molenkamp}
\affiliation{Physikalisches Institut der Universit\"{a}t
W\"{u}rzburg, 97074 W\"{u}rzburg, Germany}
\affiliation{Institute for Topological Insulators, 97074 W\"{u}rzburg, Germany}

\maketitle

\section{Density functional theory calculations}

DFT calculations were performed using the GGA+U approximation as
implemented in the Vienna Ab-initio Simulation Package (VASP) \cite{Kresse1996}. We used
the projector augmented wave method by explicitly treating 12 and 6 valence
electrons for Hg and Te, respectively, while the d electrons of Te were kept within
the core of the PBE (J. P. Perdew, K. Burke, and M. Ernzerhof) pseudopotentials. Integration over the first
Brillouin zone was done with a 16$\times$16$\times$16 Monkhorst-Pack k-mesh centered
at $\Gamma$. For all the simulations, a 600 eV plane-wave energy cut-off was used,
with spin-orbit coupling (SOC) self-consistently included. 
The ab-initio tight-binding model has been constructed by downfolding
the bulk Hamiltonian onto a localized atomic orbitals basis, made of Hg $s$
and Te $sp^3$ orbitals. The wannier90 package was used for this purpose \cite{Mostofi2008}.
The surface spectral functions in Fig. 2 have been calculated following the method described in Refs. \cite{sancho,henk}.

The usage of an on-site Coulomb interaction parameter $U=9.4$ eV and an on-site exchange interaction parameter $J=1.0$ eV within the Dudarev scheme ($U_{eff} = U-J$) follows the approach by Anversa et al. \cite{Anversa} in order to correct the energetic ordering between the $\Gamma_6$ and $\Gamma_7$ levels. This correction is also crucial to the actual momentum space separation of the Weyl nodes. In our calculations, this separation is estimated to be $\approx 0.02$ nm$^{-1}$, i.e. about four times smaller than the value originally computed by Ryan et al. \cite{Ruan} where the necessity of a correct $\Gamma_6-\Gamma_7$ ordering was overlooked.

\newpage
\section{HRXRD analysis of the investigated sample}
\begin{figure}[ht]
	\centering
		\includegraphics[width=1\columnwidth]{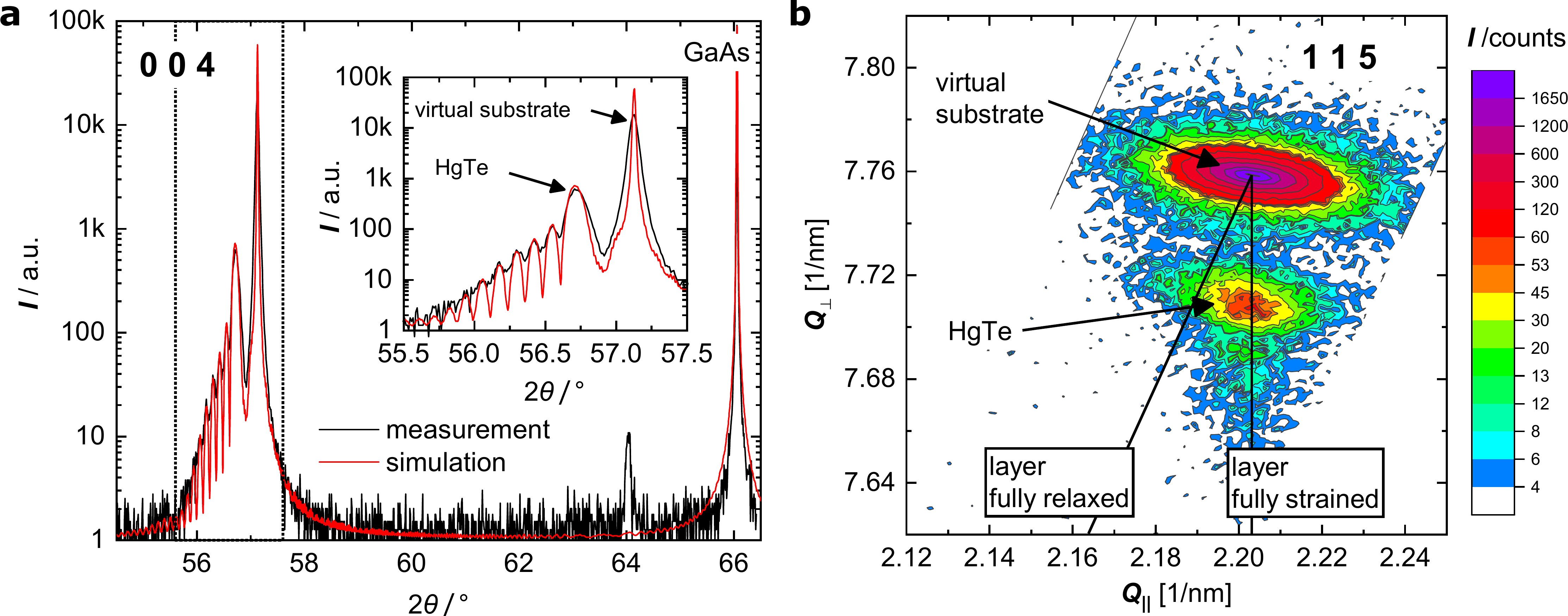}
	\caption{\textbf{a)} Measured (black) and simulated (red) diffraction pattern of the 0\,0\,4 reflection along the $2\theta$-$\omega$ direction. The inset shows a closeup around the HgTe layer peak. \textbf{b)}  Reciprocal space map of the 1\,1\,5 reflection of the superlattice virtual substrate and the HgTe layer.}
	\label{fig:XRD}
\end{figure}

To determine the degree of strain in the HgTe layer high resolution X-ray diffraction scans around the HgTe 0\,0\,4 reflection along the $2\theta$-$\omega$ direction are recorded (see Fig.~\ref{fig:XRD}\textbf{a)}) 
and analyzed following Ref.~\cite{leubner_strain_2016}. 
From the spacing between the peak of the GaAs substrate and that of the superlattice virtual substrate, we determine the out-of-plane lattice constant of our superlattice virtual substrate. 
Using the out-of-plane lattice constant and accounting for the relaxation due to the finite thickness of the superlattice, the in-plane lattice constant can be calculated. We find that the in-plane lattice mismatch between our virtual substrate and cubic HgTe is $+0.3$\%, which will induce the required compressive strain to access the Weyl-semimetal regime.

Reciprocal space maps (RSMs) of asymmetric reflections provide information about the in-plane ($Q_{\parallel}$) as well as the out-of-plane ($Q_{\perp}$) lattice constant and thus serve to determine the strain induced in our HgTe layer by the superlattice virtual substrate. 
Fig. \ref{fig:XRD}\textbf{b)} shows a RSM of the 1\,1\,5 reflection of the virtual substrate and the HgTe layer. 
The peak of the virtual substrate is used as a reference for the HgTe layer peak.  
The in-plane lattice constant given by the position $Q_{\parallel}$ is the same for both materials while the out-of-plane position $Q_{\perp}$ is different. 
In the case of (partial) relaxation, this peak would shift to the left towards the black line between the virtual substrate peak position and the origin of reciprocal space. This establishes that our HgTe layer is grown fully strained on the superlattice virtual substrate.

Combining the above results, the agreement of the HgTe peak position of simulated and measured diffraction pattern ((see insert of Fig.~\ref{fig:XRD}\textbf{a)}) confirm the strain degree of $+0.3$\% without relaxation. This agreement would not be the case for (partially) relaxed HgTe (see supplemental information of Ref.~\cite{leubner_strain_2016} for further illustrations of simulation data from strained and relaxed HgTe layers).

\begin{figure}[ht]
	\centering
		\includegraphics[width=0.5\columnwidth]{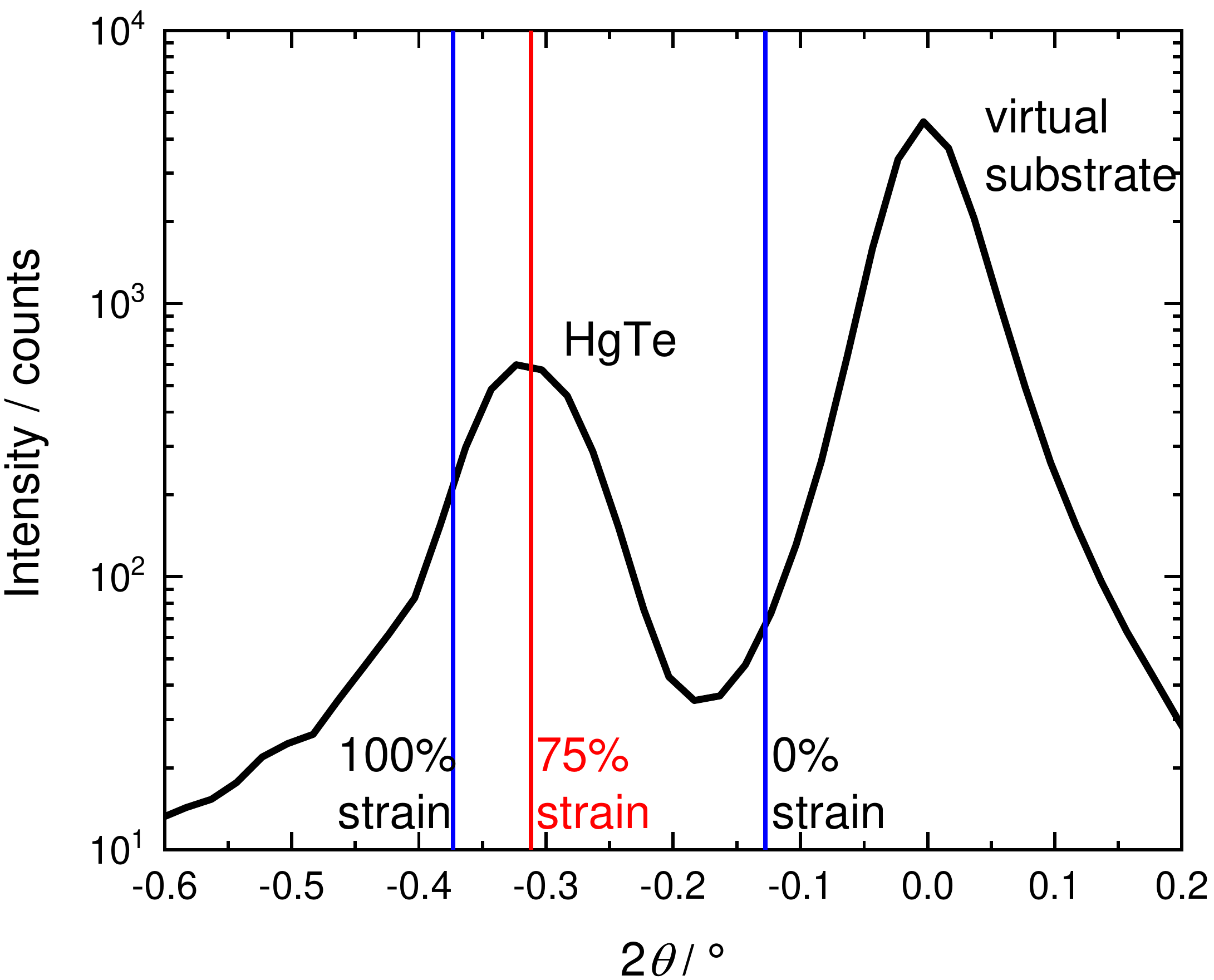}
	\caption{Measured diffraction pattern of the 0\,0\,4 reflection along the $2\theta$-$\omega$ direction with the expected peak positions for different levels of relaxation ($0\,\%, 75\,\%$ and $100\,\%$ strained) indicated.}
	\label{fig:XRD_120nm}
\end{figure}

An analogous high resolution X-ray diffraction scan around the HgTe 0\,0\,4 reflection along the $2\theta$-$\omega$ direction is recorded for the $120\,$nm sample (see Fig.~\ref{fig:XRD_120nm}).
The analysis shows that the sample (as expected for samples of this thickness) is partially relaxed by $\approx 25\,\%$.
As per the discussion in Ref.~\cite{PhysRevB.99.195423}, the partially relaxed material possess the same qualitative Weyl/Dirac band structure as fully strained material. The k-space position of the Dirac nodes as well as the energy splitting at the $\Gamma$-point are slightly reduced, which has no influence on any of our analysis.

\newpage
\section{Modeling of Volkov-Pankratov states}

To describe the experimental data provided in the main text, we perform $6\times 6$ k$\cdot$p calculations similar to those in Ref.~\cite{brune_dirac-screening_2014}. The calculations consider the compressively strained HgTe layer with hard-wall boundary conditions in the growth direction. 
We assume electrostatic decoupling of surface and bulk states. This leads to different dielectric constants at surface than in the bulk. As in Ref.~\cite{brune_dirac-screening_2014}, we find $\epsilon=21$ in the bulk and $\epsilon=3$ at the surface.

\begin{figure}[ht]
 \centering
 \includegraphics[width=1\linewidth]{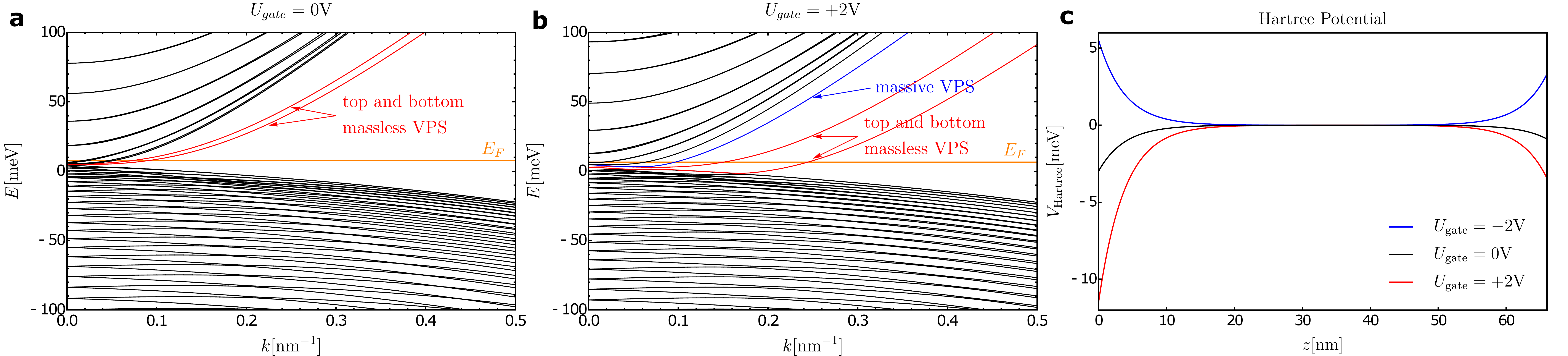}
 \caption{
Calculated k$\cdot$p band structure for a $66\mathrm{nm}$ thick sample including bulk inversion asymmetry terms and a Hartree potential corresponding to an applied gate voltage of \textbf{(a)} $U_{\text{gate}}=0\,$V ($n=2\times 10^{11}\,$cm$^{-2}$) 
  and \textbf{(b)} $U_{\text{gate}}=+2\,$V ($n=8\times 10^{11}\,$cm$^{-2}$). The bulk subbands are shown in black and the Fermi-energy is indicated by the orange line. 
  Massless Volkov-Pankratov states (VPS) are shown in red, while massive Volkov-Pankratov states are in blue.
  \textbf{c} Shape of the Hartree potential $V_{\text{Hartree}}(z)$ as a function of $z$ (growth) coordinate for gate voltages of $U_{\text{gate}}=-2\,$V (blue), $U_{\text{gate}}=0\,$V (black), and $U_{\text{gate}}=+2\,$V (red).
 }
  \label{fig:HartreeEvolution}
\end{figure}

To demonstrate the effect of gate voltage on the band structure, 
we show additional k$\cdot$p calculations for $U_{\text{gate}}=0\,$V ($n=2\times 10^{11}\,$cm$^{-2}$) and $U_{\text{gate}}=+2\,$V ($n=8\times 10^{11}\,$cm$^{-2}$) in Fig.~\ref{fig:HartreeEvolution}\textbf{a,b} using the corresponding Hartree potentials showed in Fig.~\ref{fig:HartreeEvolution}\textbf{c}. 
The asymmetry in the Hartree potential stems from the gate voltage being applied to the top of the device. 
For zero gate voltage (Fig.~\ref{fig:HartreeEvolution}\textbf{a}) the Hartree potential is to small to confine the bulk states and   therefore massive Volkov-Pankratov states cannot form. 
For a gate voltage of $+2\,$V (Fig.~\ref{fig:HartreeEvolution}\textbf{b}) the density of the massless Volkov-Pankratov states (red) at the Fermi-energy (orange) is large whereas the density of additional electron-like massive Volkov-Pankratov states (blue) is negligible.

\newpage
\section{Additional 120$\,$nm thick sample}

\begin{figure}[h]
\centering
\includegraphics[width=0.9\columnwidth]{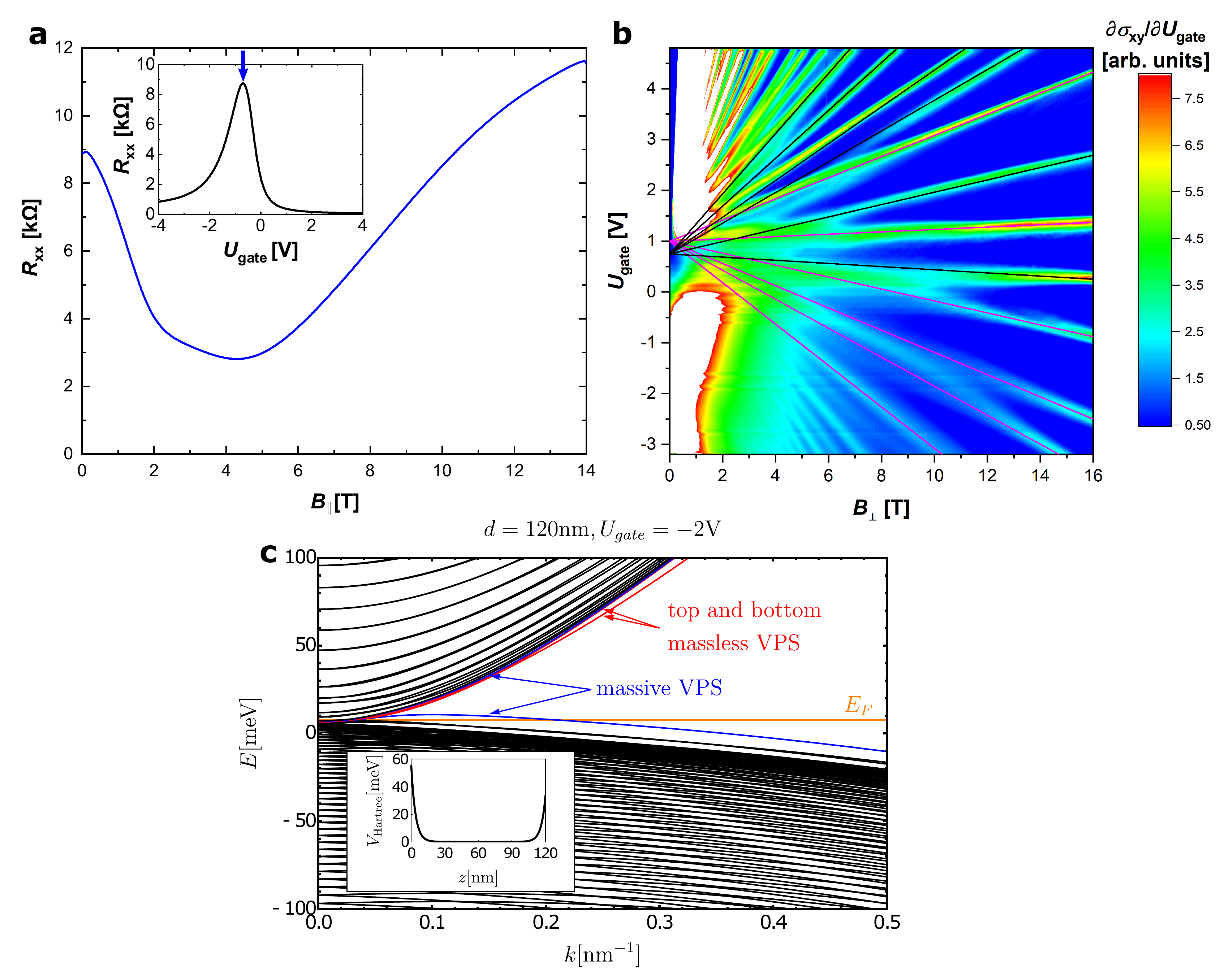}
\caption{	
\textbf{a} Longitudinal resistance of a $120\,$nm thick sample versus magnetic field $B_\parallel$ is shown for the gate voltage of the resistance maximum, indicated by an arrow in the inset.
The inset shows the longitudinal resistance $R_{xx}$ as a function of gate voltage $U_\text{gate}$.
\textbf{b} Derivative of the Hall conductance $\sigma_{xy}$ with respect to the gate voltage $U_\text{gate}$ is shown as a function of an out-of-plane magnetic field $B_\perp$ and the gate voltage $U_\text{gate}$.
The massless (massive) Volkov-Pankratov states are indicated by the black (magenta) lines.
\textbf{c} k$\cdot$p band structure including bulk asymmetry terms and a Hartree potential for a $120\,$nm thick sample and an applied gate voltage of $-2\,$V. The massless Volkov-Pankratov states are shown in red and the massive Volkov-Pankratov states in blue.
}
\label{fig:120nm_sample}
\end{figure}

To rule out any influence of the finite thickness of the sample we present here measurements and calculations on a $120\,$nm thick layer, roughly double that of the sample in the main paper, which exhibits the same transport properties.
The ability to, also in this sample, tune the Fermi level to the Weyl/Dirac node is shown in the inset to Fig.~\ref{fig:120nm_sample}\textbf{a}, where a clear maximum in the gate voltage dependence of the longitudinal resistance $R_{xx}$ is observed.
For the gate voltage corresponding to the maximum longitudinal resistance $R_{xx}$ (marked by an arrow in the inset of Fig.~\ref{fig:120nm_sample}\textbf{a}), a negative magnetoresistance is observed when the magnetic field $B_\parallel$ is aligned parallel to the current $I$.  
This is the expected behavior for the chiral anomaly. 
The derivative of the Hall conductivity $\sigma_{xy}$ with respect to the gate voltage $U_\text{gate}$ as a function of the out-of-plane magnetic field $B_\perp$ and the gate voltage $U_\text{gate}$ is shown in Fig.~\ref{fig:120nm_sample}\textbf{b}. 
Here, one can see the same qualitative Landau level dispersion as Fig.~7 from the manuscript.  Similarly to the 66nm thick sample,  the two-dimensional transport is carried by the massless Volkov-Pankratov states (indicated as black lines) and the massive Volkov-Pankratov states (magenta lines).
The corresponding k$\cdot$p calculations, which allow to identify VPS states for a $120\,$nm thick sample, are presented in Fig.~\ref{fig:120nm_sample}\textbf{c}

\bibliographystyle{apsrev4-1}
\bibliography{Supplementary}